\documentclass{webofc}
\usepackage[varg]{txfonts}   
\usepackage{natbib}
\usepackage{lineno}
\usepackage{hyperref}
\usepackage{xcolor}
\usepackage{listings}

\lstdefinestyle{customcpp}{
  belowcaptionskip=1\baselineskip,
  breaklines=true,
  language=C++,
  showstringspaces=false,
  basicstyle=\footnotesize\ttfamily,
  keywordstyle=\bfseries,
  commentstyle=\itshape,
}
\begin{document}
\title{Porting CMS Heterogeneous Pixel Reconstruction to Kokkos}

\author{\firstname{Taylor} \lastname{Childers}\inst{1}\fnsep \and
        \firstname{Matti J.} \lastname{Kortelainen}\inst{2}\fnsep\thanks{\email{matti@fnal.gov}} \and
        \firstname{Martin} \lastname{Kwok}\inst{2} \and
        \firstname{Alexei} \lastname{Strelchenko}\inst{2} \and
        \firstname{Yunsong} \lastname{Wang}\inst{3}
}

\institute{Argonne National Laboratory, Lemont, IL, USA \and
           Fermi National Accelerator Laboratory, Batavia, IL, USA \and
           Lawrence Berkeley National Laboratory, Berkeley, CA, USA
          }

\abstract{%
  Programming for a diverse set of compute accelerators in addition to the CPU is a challenge. Maintaining separate source code for each architecture would require lots of effort, and development of new algorithms would be daunting if it had to be repeated many times. Fortunately there are several portability technologies on the market such as Alpaka, Kokkos, and SYCL. These technologies aim to improve the developer productivity by making it possible to use the same source code for many different architectures. In this paper we use heterogeneous pixel reconstruction code from the CMS experiment at the CERNL LHC as a realistic use case of a GPU-targeting HEP reconstruction software, and report experience from prototyping a portable version of it using Kokkos. The development was done in a standalone program that attempts to model many of the complexities of a HEP data processing framework such as CMSSW. We also compare the achieved event processing throughput to the original CUDA code and a CPU version of it.
}
\maketitle
%
\section{Introduction}
\label{intro}

Graphics processing units (GPUs) are being used in scientific computing because of their cost and power efficiency in solving data-parallel problems. Currently each GPU vendor provides their own APIs and programming models, that also differ from the programming of the CPU. There are, however, similarities in these GPU programming models, and in many cases the code for very core pieces of algorithms can be shared between the CPU and the GPUs, but the surrounding code arranging the data and calling the algorithms has to differ. In multi-million line code bases that have many custom algorithms and have to be maintained for tens of years, such duplication of code would require significant development and maintenance effort, and be error prone to maintain.

Over several years, many technologies for fully portable code between CPUs and compute accelerators have emerged to ease the development and maintenance burden of heterogeneous applications. These technologies include C++ libraries, such as Alpaka~\cite{worpitz:2015,zenker:2016,matthes:2017}, Kokkos~\cite{edwards:2014}, and RAJA~\cite{beckingsale:2019,raja}; SYCL~\cite{sycl} that can be implemented as libraries such as triSYCL~\cite{trisycl}  and hipSYCL~\cite{hipsycl} or as specific compilers such as ComputeCpp~\cite{computecpp} by Codeplay and DPC++~\cite{dpcpp} by Intel; compiler pragma based solutions such as OpenMP~\cite{openmp} and OpenACC~\cite{openacc}; or as standard C++ itself via parallel STL~\cite{cpp20} where the compiler is solely responsible for generating necessary code for the offloading.

In this work we explore the applicability of Kokkos for portability across CPU and GPUs using the Patatrack heterogeneous pixel reconstruction workflow~\cite{bocci:2020b} from the CMS experiment~\cite{cms:detector} at the CERN LHC~\cite{Evans:2008} as a use case for a set of realistic HEP reconstruction algorithms that are able to effectively utilize a GPU. The work was done in the context of the DOE HEP Center for Computational Excellence (HEP-CCE). We look into not only the porting of the algorithms, but also the implications of integrating such an approach into a HEP data processing software.

We mimic the setup of the CMS data processing software, CMSSW~\cite{jones:2006}. CMSSW is multi-threaded~\cite{jones:2014,jones:2015,jones:2017} using the Intel Threading Building Blocks (TBB)~\cite{tbb}, and the current plan for direct same-node compute accelerators is to build code for all supported accelerators in the same release build, express all possibilities in the configuration, and decide at runtime what code exactly to run based on hardware availability~\cite{cms:Note-2021-001,bocci:2020a}. We are looking for a single-source solution that would provide portability at least across CPU and GPUs, would be relatively easy to program with by HEP physicists, would provide adequate performance on all relevant platforms, and would require the least amount of change in the CMSSW building and data processing model. It is unlikely that all these goals would be met by a single technology, and therefore it is necessary to learn the details in all these aspects to find the best compromise.

This paper is organized as follows. The technical aspects of the Patatrack pixel reconstruction are described in Section~\ref{patatrack}. A brief introduction of Kokkos is given in Section~\ref{kokkos}. The experience of porting the original CUDA application into Kokkos is reported in Section~\ref{port}. In Kokkos' nomenclature a place that runs code is called an \emph{execution space}. We have tested Serial, Threads, CUDA, and HIP execution spaces of Kokkos, and we focus on several aspects in how Kokkos would fit into a framework like CMSSW. We have measured the event processing throughput of the Kokkos version's CPU and CUDA execution spaces, and compare those to direct CPU and CUDA implementations in Section~\ref{perf}. Conclusions are given in Section~\ref{conclusions}.

\section{Patatrack Heterogeneous Pixel Reconstruction}
\label{patatrack}

\begin{figure}[tbh]
\centering
\includegraphics[width=10cm]{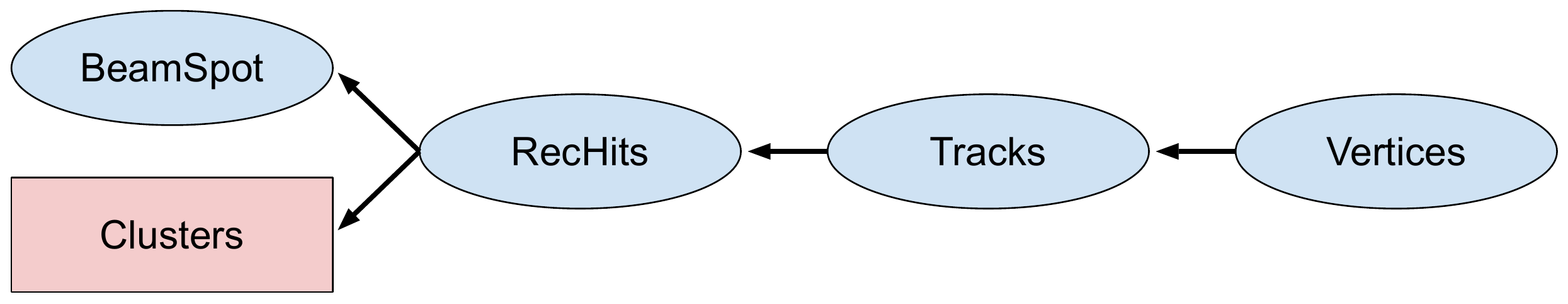}
\caption{Directed acyclic graph of the framework modules in the Patatrack pixel reconstruction. The arrows denote the data dependencies of the modules, e.g. RecHits module depends on data produced by BeamSpot and Clusters modules. The Clusters module (red rectangle) is the only one that transfer data from the device to the host and uses External Worker synchronization mechanism, while the other modules (blue oval) do not.}
\label{patatrack:modules}
\end{figure}

The Patatrack pixel reconstruction pioneered offloading algorithms to NVIDIA GPUs with direct CUDA programming within CMSSW. The offloaded chain of reconstruction algorithms takes the raw data of the CMS pixel detector as an input, along with the beamspot parameters and necessary calibration data, and produces pixel tracks and vertices. CMSSW schedules algorithms as units that are called \emph{modules}. The pixel reconstruction algorithms are organized in five modules, depicted in Figure~\ref{patatrack:modules}, that communicate the intermediate data in the GPU memory through the CMSSW event data. The BeamSpot module only transfers the beamspot data into the GPU. The Clusters module transfers the raw data to the GPU, unpacks them, calibrates the individual pixels, and clusters the pixels on each detector module. The RecHits module estimates the 3D position of each cluster forming hits. The Tracks module forms n-tuplets from the hits, and fits the hit n-tuplets to obtain track parameters. The Vertices module forms vertices from the pixel tracks. There are further modules to optionally transfer the tracks and vertices to the CPU, and to convert the Structure-of-Array (SoA) data structures to the data formats used by downstream algorithms in CMSSW, but those are not considered in this work and therefore not shown in Figure~\ref{patatrack:modules}.

In order to explore code portability technologies, the CUDA code of the Patatrack pixel reconstruction was extracted from CMSSW into a standalone program~\cite{standalone}. The separation from CMSSW gives us freedom to modify the compilers, build rules, external libraries, and code organization that would be more laborious to achieve in the full CMSSW software stack. The standalone program was crafted to mimic several aspects of CMSSW, including similar organization of code into shared libraries, plugin libraries that are loaded dynamically based on run-time information, and a simple framework that uses TBB for multi-threading. From the CMSSW framework concurrency features this simple framework includes only event loop based on TBB tasks, processing of multiple events concurrently, and processing of independent modules concurrently for the same event. There is only a single module type of each module having a separate instance for each concurrent event, and the \textit{External Worker} concept~\cite{bocci:2020a} is included in order to use the CPU threads to do other work while the GPU is running the offloaded work. The CMSSW tools to use CUDA runtime directly in the modules~\cite{bocci:2020a} are also included.

The standalone setup includes a binary data file that contains raw pixel detector data from 1000 simulated top quark pair production events from CMS Open Data~\cite{ttbardata}, with an average of 50 superimposed pileup collisions with a center-of-mass energy of $13\,\textrm{TeV}$, using design conditions corresponding to the 2018 CMS detector. All of the data, about $250\,\textrm{MB}$, are read into the memory at the job startup to exclude I/O from the throughput measurement. The necessary pixel detector conditions data are also stored in binary files, and read into the memory at the start of the job. The data processing throughput is calculated by measuring the time spent in the event processing, and dividing the number of processed events with that time. For each event, the object holding the raw data for that event is copied once from the aforementioned memory buffer to another object owned by the event data structure. The event processing time includes the time taken by this copy operation.

\section{Kokkos}
\label{kokkos}

Kokkos is a programming model and a C++ library for writing performance portable applications. At the time of writing the latest version of Kokkos is 3.3.1, and it supports several execution spaces. An algorithm can be run serially on the host CPU via a \emph{host serial} execution space, or it can be parallelized with one of two \emph{host parallel} execution spaces that are OpenMP and (POSIX) Threads. An algorithm can also be offloaded to compute accelerators with \emph{device parallel} execution spaces. NVIDIA GPUs can be used with CUDA or HPX execution spaces, and AMD GPUs can be used with HIP execution space. There are also OpenMP-Target and SYCL 2020 execution spaces that can support various platforms depending on the underlying toolchain. Currently all other device parallel execution spaces than CUDA are experimental. In this work we have tested Serial, Threads, CUDA, and HIP execution spaces.

Kokkos makes use of a runtime library. The library can have the Serial, one host parallel, and one device parallel execution space enabled at the same time, and this set is chosen at the library build configuration time. In addition, at least for CUDA execution space, one library can support only GPUs that have the same major compute capability number. For example, one library can support Volta (compute capability 7.0) and Turing (7.5) GPUs, but the same library can not support Ampere (8.0) or Pascal (6.0) GPUs. In the code the execution space to be used can be chosen at compile time with template arguments. If the execution space is not specified explicitly, the most advanced execution space available in the library is used, i.e. device parallel execution space is preferred over host parallel execution space, which is preferred over the Serial execution space. Currently Kokkos supports only one device per process.

Kokkos provides high-level interface for parallel operations. These include \texttt{parallel\_for} for a for-loop of independent iterations, \texttt{parallel\_scan} for a prefix scan, and \texttt{parallel\_reduce} for a reduction. Parallel operations can be nested with some restrictions. The details of the iteration are controlled with a policy. A \texttt{RangePolicy} can be used for a 1-dimensional range where all elements of the range can be processed independently. An example of \texttt{parallel\_for} with \texttt{RangePolicy} is shown in Figure~\ref{fig:kokkosRangePolicy} and a corresponding CUDA version in Figure~\ref{fig:kokkosRangePolicyCUDA}. An \texttt{MDRangePolicy} extendes the concept of the 1-dimensional \texttt{RangePolicy} to many, up to 6, dimensions. A \texttt{TeamPolicy} introduces a \emph{league} of \textit{teams} that consist of \emph{threads}\footnote{The \emph{league} corresponds to \emph{grid} in CUDA, and \emph{team} corresponds to \emph{block}.}. Threads in a team can use a common scratch memory space, and can synchronize within the team with a barrier. In addition, Kokkos has some support for tasks and graphs, that are not explored in this work.

\begin{figure}[btp]
\begin{lstlisting}[language=C++,style=customcpp]
// declarations of variables
constexpr uint32_t MaxNumModules;
constexpr uint32_t maxHitsInModule();
Kokkos::View<uint32_t const*, Kokkos::CudaSpace> cluStart;
Kokkos::View<uint32_t*, Kokkos::CudaSpace> moduleStart;

Kokkos::parallel_for(
  Kokkos::RangePolicy<Kokkos::Cuda>(0, MaxNumModules)),
  KOKKOS_LAMBDA(const int index) {
    moduleStart(index + 1) = std::min(maxHitsInModule(), cluStart(index));
});
\end{lstlisting}
\caption{A simplified example of using \texttt{RangePolicy} policy with \texttt{parallel\_for}. The initialization of the declared variables is omitted for brevity. In this example the execution and memory space template argument are spelled out explicitly. If the compile-time defaults for those suffice, the explicit template arguments can be left out.  Corresponding CUDA program is shown in Figure~\ref{fig:kokkosRangePolicyCUDA}.}
\label{fig:kokkosRangePolicy}
\end{figure}

\begin{figure}[tbp]
\begin{lstlisting}[language=C++,style=customcpp]
// declarations of used variables
constexpr uint32_t MaxNumModules;
constexpr uint32_t maxHitsInModule();

__global__
void fillHitsModuleStart(uint32_t const* cluStart, uint32_t* moduleStart) {
  for(int i = threadIdx.x, iend = MaxNumModules; i < iend; i += blockDim.x) {
    moduleStart[i + 1] = std::min(maxHitsInModule(), cluStart[i]);
  }
}

uint32_t const* cluStart_;
uint32_t* moduleStart_;
fillHitsModuleStart<<<1, 1024>>>(cluStart_, moduleStart_);
\end{lstlisting}
\caption{CUDA version of the simplified example expressed in Kokkos in Figure~\ref{fig:kokkosRangePolicy}. The initialization of the declared variables is omitted for brevity.}
\label{fig:kokkosRangePolicyCUDA}
\end{figure}

As well as parallel operations, Kokkos provides a datastructure for multi-dimensional array, \texttt{Kokkos::View}. It is reference counted and behaves like \texttt{std::shared\_ptr}, and can be passed to device functions by value. A major feature of the \texttt{Kokkos::View} is that its memory layout can be controlled with template arguments, and  the default layout depends on the memory space. In addition, intents for the memory can be expressed with additional template arguments, for example specifying random-access constant data enables seamless use of CUDA texture caches. Data transfers between the host and the device are done explicitly.

\section{Porting experience}
\label{port}

\subsection{Impact on building}

The current plan to support compute accelerators in CMSSW software stack is to build code for all supported accelerators, and choose the exact version to be run at runtime~\cite{cms:Note-2021-001}. The various constraints of the Kokkos runtime library, described in Section~\ref{kokkos}, make it challenging to deploy in this manner. A single runtime library supporting only one device parallel execution space, and only one CUDA major architecture or CPU vector architecture, would, in this plan, imply the need to build many versions of the runtime library. The correct version would have to be loaded dynamically based on the available hardware. In this work we used exactly one runtime library at a time.

Every source file that includes any Kokkos header must be built with a compiler that is capable of compiling the code for all the enabled execution spaces, even if the source file would not use any Kokkos functionality. For example, if the Kokkos runtime library was built with CUDA execution space enabled, all source files including Kokkos headers must be compiled with a CUDA capable compiler. 

Kokkos provides an integration with the CMake build system. In this work, however, we used CMake only to build the Kokkos runtime library itself, and we used a plain Makefile to build the application code. We did this because CMSSW uses the SCRAM build system~\cite{wellisch:2003}, and therefore we'd have to understand the exact build rules in order to implement those for SCRAM. 

The inability of \texttt{nvcc} to link device code from shared objects imposed severe constraints on how the Kokkos runtime library had to be built. We were able to use the runtime library built as a dynamic library with \texttt{RangePolicy}, but with the first use of \texttt{TeamPolicy} that approach lead to link errors from \texttt{nvcc}. The only build setup we managed to get to work was to build the Kokkos runtime library as a static library without support for relocatable device code, but with position-independent code for the host (\texttt{-fPIC}) to be able to link the static library with dynamic libraries of the application. This setup implies that CUDA separate compilation model can not be used, and therefore each source file must contain all device code called from that file, either directly or via including other files. Also, CUDA dynamic parallelism can not be used.

With the HIP execuion space we were able to use a dynamic Kokkos runtime library, and in fact were not able to get a static build to work with the HIP compiler.

\subsection{Impact on code}

As mentioned in Section~\ref{kokkos}, the Kokkos execution space is chosen at compile time. A choice done at runtime would be a much better fit in the current plans of using compute accelerators in CMSSW. We implemented the capability of choosing the execution space at runtime by building each source file containing Kokkos code once for each execution space and using namespaces to guarantee different symbols for each execution space.

Conversion of CUDA kernel calls to Kokkos parallel operations was mostly straightforward. Kokkos provides a parallel scan and sort, and therefore we decided to use those instead of trying to port the implementations of scan and radix sort device functions in the direct CUDA version. The code uses team-wide scan, but before version 3.3, Kokkos provided only league-wide scan. Before updating to Kokkos 3.3 we used the league-wide scan with two additional kernels to post-process the league-wide result to be equivalent to a team-wide scan. Kokkos' parallel sort function can be called only from the host code, which meant that we had to split all the CUDA kernels that called the device-side sort function into two kernels, and call the Kokkos' host-side sort function in between. Finding out the proper and efficient way to transform the CUDA code to use the Kokkos' scan and sort APIs was the most time consuming single effort.

For hierarchical parallelism, or thread teams, we found that the number of threads in a team is not exactly portable. The Serial execution space requires it to be exactly one, Threads execution space can use at most the number of CPU threads, and CUDA execution space has the same limitations as CUDA itself. This disparity can be largely mitigated by specifying the number of threads as \texttt{Kokkos::AUTO()}, that leaves the decision of the number of threads to Kokkos.

We found \texttt{Kokkos::View} to be useful by providing a unified interface for memory allocation, and smart pointer semantics for managing the ownership of the memory block. Also the ability to avoid an additional memory allocation in code that transfers data from host to device for CPU-only execution spaces is a plus. The more advanced features like multiple dimensions and the layout control are not needed in this code, where nearly all arrays have only one dimension. The only exception is the track covariance matrix, but we did not try to transform the Eigen-based implementation in the original CUDA into multidimensional \texttt{Kokkos::View}. In this code a SoA abstraction would be much more useful than multi-dimensional array, and we do not see how \texttt{Kokkos::View} would help in crafting SoA data structures.

In the first Kokkos version we found that about $80\,\%$ overall kernel runtime was spent in \texttt{Kokkos::View} initialization. In this code the first operation for all device memory is a write either by a memory copy from the host memory, or by a computation done in a kernel. Therefore all the initialization done by default is unnecessary, and avoiding that with \texttt{Kokkos::ViewAllocateWithoutInitializing} argument to \texttt{Kokkos::View} constructor improved the event processing throughput by almost a factor of 3.

At the time of writing, we have not been able to successfully run the full application with the HIP execution space. A test application that uses the same build and dynamic library infrastructure works well, but is not complex-enough to give meaningful insights into the performance.

Furthermore, we have not yet managed to run the application with multiple concurrent events with Serial or CUDA execution spaces. The Threads execution space explicitly prevents calls from more than one thread, even if the calls would come at different times. Despite of the Threads execution space being uninteresting to be used in the context of CMSSW, we have included it as a comparison point in the performance measurements in Section~\ref{perf} to show how a parallelization strategy different from concurrent events would perform.

\section{Performance comparison}
\label{perf}

The performance tests were done on GPU nodes of the Cori supercomputer at the National Energy Research Scientific Computing Center (NERSC). A Cori GPU node has two sockets with Intel Xeon Gold 6148 ("Skylake") processors, each with 20 cores and 2 threads per core, and eight NVIDIA V100 GPUs. For this work we used only one CPU socket, to avoid the impact of non-uniform memory access (NUMA), and one GPU. In all tests the threads were pinned to a single socket. Each job was run for approximately 5 minutes, processing the set of 1000 individual events for an integer number of times, and repeated 8 times on random nodes of the GPU cluster. The code was compiled with GCC 8.3.0, and \texttt{nvcc} from CUDA 11.1.

In order to minimize the impact of the CPU frequency scaling the CPU programs were tested by running another program on the background with as many threads as needed to fill all the 40 hardware threads of the socket. Table~\ref{perf:cpu} shows the throughput of the Kokkos version with Serial and Threads execution spaces, and of the direct CPU version with 1 and 40 threads. The Kokkos version processes one event at a time, and with the Threads execution space each Kokkos parallel operation is parallelized with the same number of threads. The direct CPU version, on the other hand, is parallelized by processing multiple events concurrently, one event per thread. While comparing the multi-threaded throughput of these two approaches is not exactly fair, it does show what can be achieved with a single process using the different approaches.

The results in Table~\ref{perf:cpu} show that the intra-event parallelization scales poorly, whereas parallelizing over events gives much better throughput and scales well. We have not concluded yet why the direct CPU version gives 1.5 times better throughput than the Kokkos version with Serial execution space.

\begin{table}[b]
\centering
\caption{Comparison of the event processing throughput between the Kokkos version of the program using Serial and Threads execution spaces and the CPU version implemented from the original CUDA version through a simple translation header. In all cases all the threads were pinned to a single CPU socket (Intel Xeon Gold 6148) that has 20 cores and 2 threads per core. Each test ran about 5 minutes, and CPU-heavy threads from a background process were used to fill all the 40 hardware threads of the socket. The work in the CPU version is parallelized by processing as many events concurrently as the number of threads the job uses without any intra-event parallelization, whereas in the Kokkos version there is only one event in flight, and all parallelization is within the data of that event. For the Kokkos version with Threads execution space the maximum throughput from a scan from 1 to 20 threads is reported. The reported uncertainty corresponds to sample standard deviation of 8 trials.}
\label{perf:cpu}       
\begin{tabular}{lc}
\hline
Test case & Throughput (events/s) \\
\hline
CPU version, 1 thread & $13.5 \pm 0.2$ \\
Kokkos version, Serial execution space & $8.5 \pm 0.2$ \\
\hline
CPU version, 40 threads & $539\pm 9$ \\
Kokkos version, Threads execution space, peak (18 threads) & $28 \pm 1$ \\
\hline
\end{tabular}
\end{table}

The programs using CUDA were tested without any background activity on the CPU. Table~\ref{perf:cuda} shows the throughput of the Kokkos version with CUDA execution space, and of the direct CUDA version. The direct CUDA version can process data from multiple events concurrent with CUDA streams, and this approach helps to get 2.5 times higher throughput from the V100 GPU than when processing one event at a time. With a single event in flight, the memory pool, based on the \texttt{CachingDeviceAllocator} of the CUB~\cite{cub} library, helps to increase the throughput by 4.5 times compared to using raw CUDA memory allocations.

The Kokkos version with the CUDA execution space reaches about $70\,\%$ of the throughput of the direct CUDA version when run on a single concurrent event and disabling the use of the memory pool. Profiling indicates that various overheads e.g. in the \texttt{Kokkos::View} are the main cause for the performance difference. From Table~\ref{perf:cuda} it is also clear that the kind of data processing done in this application benefits greatly from a memory pool, and from processing multiple events concurrently.

\begin{table}[tb]
\centering
\caption{Comparison of the event processing throughput between the Kokkos version of the program using CUDA execution space and the original CUDA version. In all cases the CPU threads were pinned to a single CPU socket, and used one NVIDIA V100 GPU. Each test ran about 5 minutes, and the machine was free from other activity. The CUDA version can process data from multiple events concurrently using many CPU threads and CUDA streams, and uses a memory pool to amortize the cost of raw CUDA memory allocations. The maximum throughput from a scan from 1 to 20 concurrent events is reported for the CUDA version. In order to compare to the current state of the Kokkos version, the CUDA version was tested also with 1 concurrent event and disabling the use of the memory pool. The reported uncertainty corresponds to sample standard deviation of 8 trials.}
\label{perf:cuda}       
\begin{tabular}{lc}
\hline
Test case & Throughput (events/s) \\
\hline
CUDA version, peak (9 concurrent events and CPU threads) & $1840\pm 20$ \\
CUDA version, 1 concurrent event & $720 \pm 20$ \\
CUDA version, 1 concurrent event, memory pool disabled & $159 \pm 1$ \\
\hline
Kokkos version, CUDA execution space & $115.7 \pm 0.3$ \\
\hline
\end{tabular}
\end{table}

\section{Conclusions}
\label{conclusions}

We have ported the Patatrack heterogeneous pixel reconstruction code from CUDA to Kokkos. In our experience Kokkos provides an API that is at a higher level than CUDA, and would be easier to develop new algorithms by physicists that are not necessarily experts in programming. We have achieved almost full portability between CPU, CUDA, and HIP, even if work still continues to understand runtime failures of the HIP execution space version of the code.
This analysis shows that Kokkos can give $70\,\%$ of native CUDA performance in a simplified setup without either a memory pool or concurrent events. If similar performance proportion can be achieved also in a more realistic setup, it may be worth using a portable framework to reduce person power in maintaining a code base despite the loss of compute performance.

Our impression is that Kokkos would work well for a project that compiles the code separately for each target architecture, does not rely much on shared libraries, uses CMake as the build system, and does not rely on concurrent work outside of Kokkos. CMSSW doing all these in the opposite way implies that integrating the current version of Kokkos into the current data processing model of CMSSW would be challenging to do without sacrificing application performance. It is not, however, clear to us at this time to what extent these challenges are caused by design choices in Kokkos, or by the nature of the portability problem itself.

More work is needed to complete the study with Kokkos. In addition, comparisons to other portability technologies are planned within the HEP-CCE.

\section*{Acknowledgements}

This work was supported by the U.S. Department of Energy, Office of
Science, Office of High Energy Physics, High Energy Physics Center for
Computational Excellence (HEP-CCE) at Argonne National Laboratory,
Fermi National Accelerator Laboratory, and Lawrence Berkeley National
Laboratory under B\&R KA2401045. This research used resources of the
National Energy Research Scientific Computing Center (NERSC), a U.S.
Department of Energy Office of Science User Facility located at
Lawrence Berkeley National Laboratory, operated under Contract No. DE-
AC02-05CH11231.

\bibliography{references}

\begin{thebibliography}{28}

\bibitem{worpitz:2015}
B.~Worpitz, \emph{Investigating performance portability of a highly scalable
  particle-in-cell simulation code on various multi-core architectures} (2015)

\bibitem{zenker:2016}
E.~Zenker, B.~Worpitz, R.~Widera, A.~Huebl, G.~Juckeland, A.~Kn{\"{u}}pfer,
  W.E. Nagel, M.~Bussmann, \emph{Alpaka - An Abstraction Library for Parallel
  Kernel Acceleration} (IEEE Computer Society, 2016), \texttt{1602.08477}

\bibitem{matthes:2017}
A.~{Matthes}, R.~{Widera}, E.~{Zenker}, B.~{Worpitz}, A.~{Huebl},
  M.~{Bussmann}, \emph{Tuning and optimization for a variety of many-core
  architectures without changing a single line of implementation code using the
  Alpaka library} (2017), \texttt{1706.10086}

\bibitem{edwards:2014}
H.C. Edwards, C.R. Trott, D.~Sunderland, Journal of Parallel and Distributed
  Computing \textbf{74}, 3202 (2014)

\bibitem{beckingsale:2019}
D.A. Beckingsale, J.~Burmark, R.~Hornung, H.~Jones, W.~Killian, A.J. Kunen,
  O.~Pearce, P.~Robinson, B.S. Ryujin, T.R.W. Scogland, \emph{RAJA: Portable
  Performance for Large-Scale Scientific Applications} (2019), IEEE/ACM
  International Workshop on Performance, Portability and Productivity in HPC
  (P3HPC), p.~71

\bibitem{raja}
\emph{{RAJA Performance Portability Layer}}, \url{https://github.com/LLNL/RAJA}
  (2021), accessed: 2021-02-07

\bibitem{sycl}
{The Khoronos SYCL Working Group}, \emph{{SYCL} 2020 Specification (revision
  2)} (2021)

\bibitem{trisycl}
\emph{trisycl}, \url{https://github.com/triSYCL/triSYCL} (2021), accessed:
  2021-02-07

\bibitem{hipsycl}
A.~Alpay, V.~Heuveline, \emph{SYCL beyond OpenCL: The Architecture, Current
  State and Future Direction of HipSYCL}, in \emph{Proceedings of the
  International Workshop on OpenCL} (Association for Computing Machinery, New
  York, NY, USA, 2020), IWOCL '20,
  \urlstyle{tt}\url{https://doi.org/10.1145/3388333.3388658}

\bibitem{computecpp}
\emph{{ComputeCpp}},
  \url{https://developer.codeplay.com/products/computecpp/ce/home} (2021),
  accessed: 2021-02-07

\bibitem{dpcpp}
\emph{Intel {oneAPI DPC++/C++} compiler},
  \url{https://software.intel.com/content/www/us/en/develop/tools/oneapi/components/dpc-compiler.html}
  (2021), accessed: 2021-02-07

\bibitem{openmp}
{OpenMP Architecture Review Board}, \emph{OpenMP Application Programming
  Interface, version 5.1} (2020)

\bibitem{openacc}
{OpenACC-Stadnrad.org}, \emph{The OpenACC Application Programming Interface,
  version 3.1} (2020)

\bibitem{cpp20}
ISO/IEC 14882:2020, \emph{Programming languages - {C++}} (2020)

\bibitem{bocci:2020b}
A.~Bocci, V.~Innocente, M.~Kortelainen, F.~Pantaleo, M.~Rovere, Front. Big.
  Data \textbf{3}, 601728 (2020), \texttt{2008.13461}

\bibitem{cms:detector}
{CMS Collaboration}, JINST \textbf{3}, S08004 (2008)

\bibitem{Evans:2008}
L.~Evans, P.~Bryant, JINST \textbf{3}, S08001 (2008)

\bibitem{jones:2006}
C.D. Jones, M.~Paterno, J.~Kowalkowski, L.~Sexton-Kennedy, W.~Tanenbaum,
  \emph{The New {CMS} Event Data Model and Framework}, in \emph{Proceedings of
  International Conference on Computing in High Energy and Nuclear Physics
  (CHEP06)} (2006)

\bibitem{jones:2014}
C.D. Jones, E.~Sexton-Kennedy, J. Phys.: Conf. Series \textbf{513}, 022034
  (2014)

\bibitem{jones:2015}
C.D. Jones, L.~Contreras, P.~Gartung, D.~Hufnagel, L.~Sexton-Kennedy, J. Phys.:
  Conf. Series \textbf{664}, 072026 (2015)

\bibitem{jones:2017}
C.D. Jones, J. Phys.: Conf. Series \textbf{898}, 042008 (2017)

\bibitem{tbb}
\emph{{oneAPI Threading Building Blocks}},
  \url{https://github.com/oneapi-src/oneTBB} (2021), accessed: 2021-02-07

\bibitem{cms:Note-2021-001}
{CMS Offline Software and Computing}, \emph{Evolution of the {CMS} computing
  model towards phase-2} (2021), {CMS-NOTE-2021-001},
  \urlstyle{tt}\url{https://cds.cern.ch/record/2751565}

\bibitem{bocci:2020a}
A.~Bocci, D.~Dagenhart, V.~Innocente, C.~Jones, M.~Kortelainen, F.~Pantaleo,
  M.~Rovere, EPJ Web Conf. \textbf{245}, 05009 (2020)

\bibitem{standalone}
\emph{Standalone {Patatrack} pixel tracking},
  \url{https://github.com/cms-patatrack/pixeltrack-standalone/} (2021),
  accessed: 2021-02-07

\bibitem{ttbardata}
{CMS Collaboration}, \emph{{TTToHadronic\_TuneCP5\_13TeV-powheg-pythia8 in
  FEVTDEBUGHLT format for 2018 collision data. CERN Open Data Portal.}},
  \href{http://doi.org/10.7483/OPENDATA.CMS.GOB0.0LEW}{doi:10.7483/OPENDATA.CMS.GOB0.0LEW}
  (2019)

\bibitem{wellisch:2003}
J.P. Wellisch, C.~Williams, S.~Ashby, \emph{{SCRAM}: Software configuration and
  management for the {LHC} Computing Grid project}, in \emph{Proceedings of
  International Conference on Computing in High Energy and Nuclear Physics
  (CHEP03)} (2003), p. TUJP001, \texttt{cs/0306014}

\bibitem{cub}
\emph{{CUB}}, \url{https://nvlabs.github.io/cub/} (2021), accessed: 2021-02-07

\end{thebibliography}

\end{document}